\documentclass[11pt]{article}
\usepackage{amsmath}
\usepackage{graphicx}
\usepackage{array}
\usepackage{booktabs}
\usepackage{subfigure}
\usepackage{float}
\usepackage{booktabs,makecell,multirow}
\usepackage{url}
\usepackage{color}

\newcommand{\eqnb}{\begin{equation}}
\newcommand{\eqne}{\end{equation}}

\begin{document}

\title{Performance Evaluation, Optimization and Dynamic Decision in
Blockchain Systems: A Recent Overview}
\author{ Quan-Lin Li$^1$, Yan-Xia Chang$^1$\thanks{Corresponding author: Y. X. Chang (changyanxia@emails.bjut.edu.cn)\newline ----- Tell us your papers to add to the next update} and Qing Wang$^2$ \\
$^1$School of Economics and Management\\
Beijing University of Technology, Beijing 100124, China \\
$^2$Monash Business School, Monash University \\
900 Dandenong Road, Caulfield East, VIC, 3145, Australia.}
\maketitle

\begin{abstract}
With rapid development of blockchain technology as well as integration of
various application areas, performance evaluation, performance optimization, and dynamic
decision in blockchain systems are playing an increasingly important role in
developing new blockchain technology. This paper provides a recent
systematic overview of this class of research, and especially, developing
mathematical modeling and basic theory of blockchain systems. Important
examples include (a) performance evaluation: Markov processes, queuing
theory, Markov reward processes, random walks, fluid and diffusion
approximations, and martingale theory; (b) performance optimization: Linear
programming, nonlinear programming, integer programming, and multi-objective programming; (c) optimal
control and dynamic decision: Markov decision processes, and stochastic
optimal control; and (d) artificial intelligence: Machine learning, deep
reinforcement learning, and federated learning. So far, a little research
has focused on these research lines. We believe that the basic theory with
mathematical methods, algorithms and simulations of blockchain systems
discussed in this paper will strongly support future development and
continuous innovation of blockchain technology.

\end{abstract}

\section{Introduction}

Since Bitcoin was proposed by Nakamoto \cite{Nak:2008} in 2008, blockchain
technology has received tremendous attention from both practitioners and
academics. So far, blockchain has made remarkable progress by means of many
interesting and creative combinations of multiple key computer technologies, such as
distributed systems, consensus mechanism, network and information security,
privacy protection, encryption technology, peer-to-peer networks, edge
computing, Internet of Things, and artificial intelligence. At the
same time, some effective scalable frameworks and security designs of
blockchain have been further developed, for example, off-chain, side-chain,
cross-chain, shard, fault tolerant, and attack detection. However,
compared with rapid development of blockchain technology, mathematical
modeling and analysis of blockchain systems is relatively backward, thus it is clear that developing blockchain technology
extremely needs such important basic theory and necessary mathematical
methods.

In this paper, we review mathematical modeling and analysis methods in some
aspects (but no completeness) of blockchain technology, including some
important progress that can further drive developing new potential blockchain
technologies. To this end, our overview in this paper is listed as follows:
\textbf{(1)} Mining processes and management; \textbf{(2)} consensus
mechanism; \textbf{(3)} performance evaluation; \textbf{(4)} performance
optimization; \textbf{(5)} optimal control and dynamic decision; \textbf{(6)}
machine learning; \textbf{(7)} blockchain economy and market; and \textbf{(8)%
} blockchain ecology. Note that the eight survey points aim to setting up
stochastic models and associated mathematical methods to theoretically improve
blockchain's performance, scalability, security, privacy protection, work
efficiency, and economic benefit. In what follows, we use Figures 1 to 8 to
describe and analyze the eight survey points \textbf{(1) }to\textbf{\ (8) }simply.

\textbf{(1) Mining processes and management}

For mathematical modeling and analysis on this research direction, we need
to discuss the key system factors or parameters that largely influence performance, scalability, security, and privacy protection of blockchain systems. For example, the miners,
the mining pools, the difficulty of solving the cryptographic puzzle, the
transaction fee, the blockchain reward, the competitive behavior, the tree
with forked structure, the work efficiency, the economic benefit; the attack
strategies, the security, the vulnerability, the fault tolerance, and privacy protection. See Figure
1 for more details.

\begin{figure*}[tbph]
\centering                 \includegraphics[width=10cm]{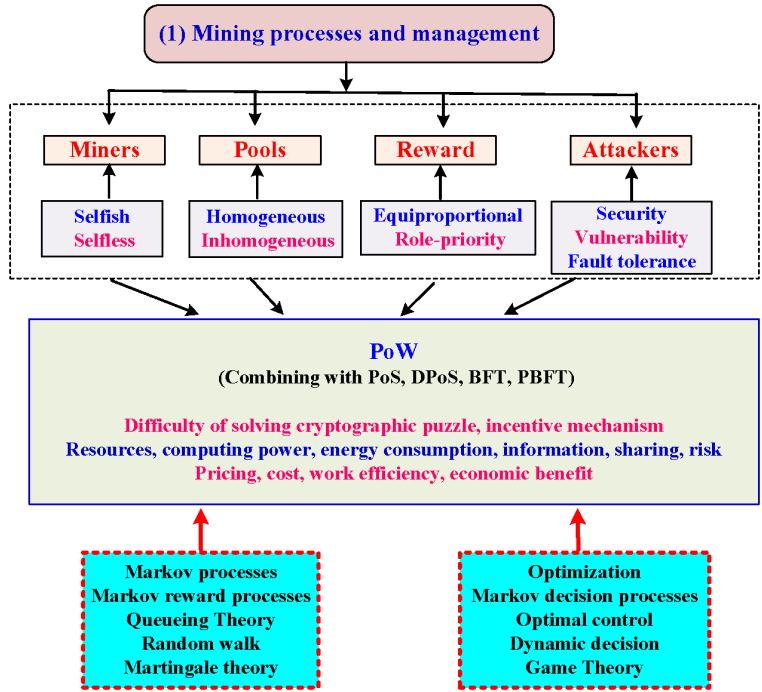} \centering
\caption{The mining processes and management}
\label{Fig-1}
\end{figure*}

\textbf{(2) Consensus mechanism}

For mathematical modeling and analysis on this research direction, we need
to discuss the random consensus-accomplished times for different consensus
protocols (or algorithms), such as PoW, PoS, DPoS, BFT, PBFT, and Raft. Furthermore, we need to analyze the blockchain systems under
different consensus protocols and to study the throughput, security, privacy
protection, and scalability of blockchain systems. Our main concerns include
a set of basic factors, such as consensus types, efficiency, convergence,
consistency, network delay, and energy consumption. See Figure 2 for
more details.

\begin{figure*}[tbph]
\centering                 \includegraphics[width=10cm]{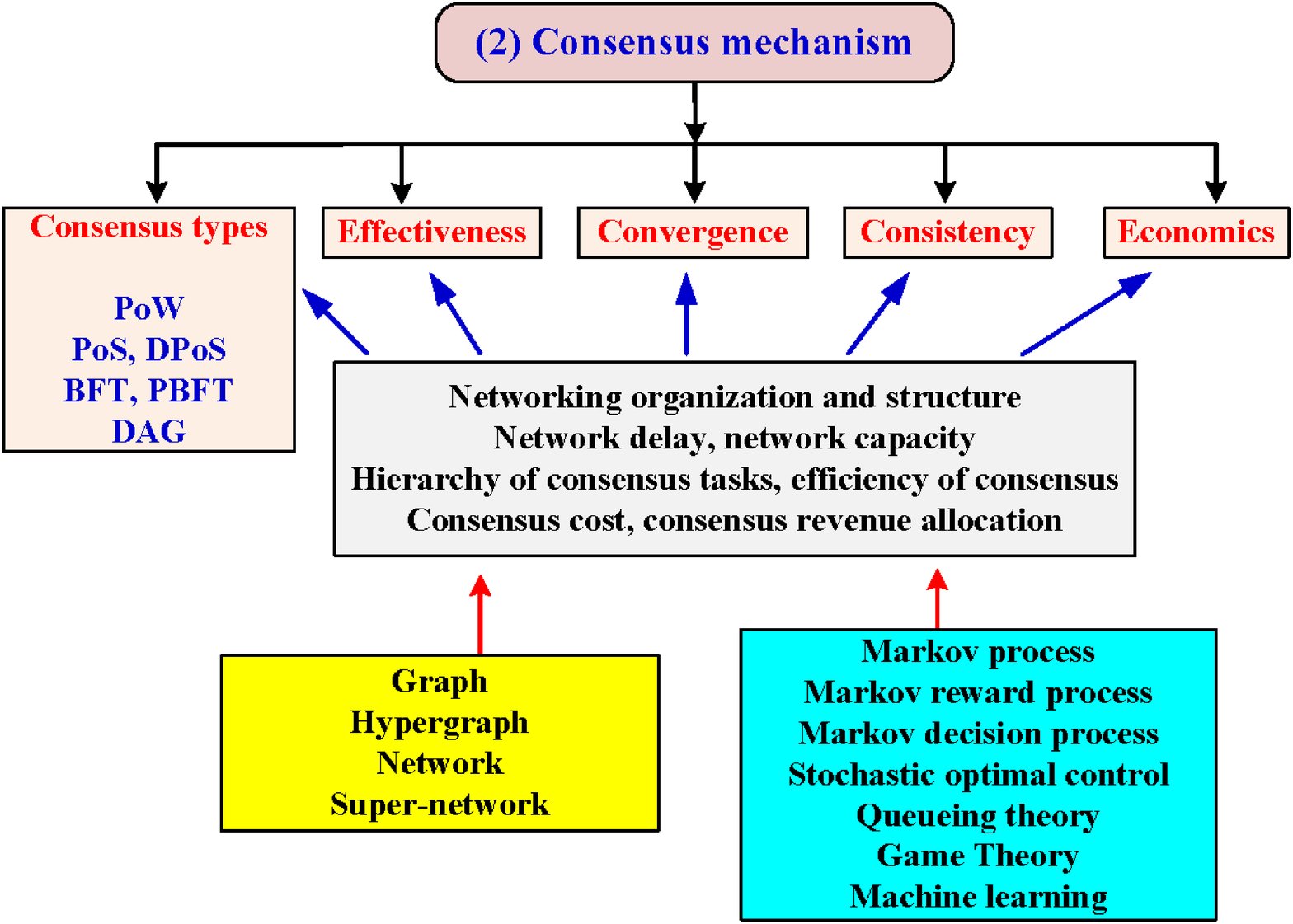} \centering
\caption{The consensus mechanism}
\label{Fig-2}
\end{figure*}

\textbf{(3) Performance evaluation}

In this class of mathematical modeling and analysis, we need to set up
performance models of blockchain systems when considering different
consensus mechanisms or protocols or algorithms (PoW, PoS, DPoS, PBFT, DAG
and so on), different blockchain types (Bitcoin, Ethereum, side-chain,
cross-chain, off-chain and so on), and innovation and new network architectures of
blockchain systems. See Figure 3 for more details.

\begin{figure*}[tbph]
\centering                 \includegraphics[width=10cm]{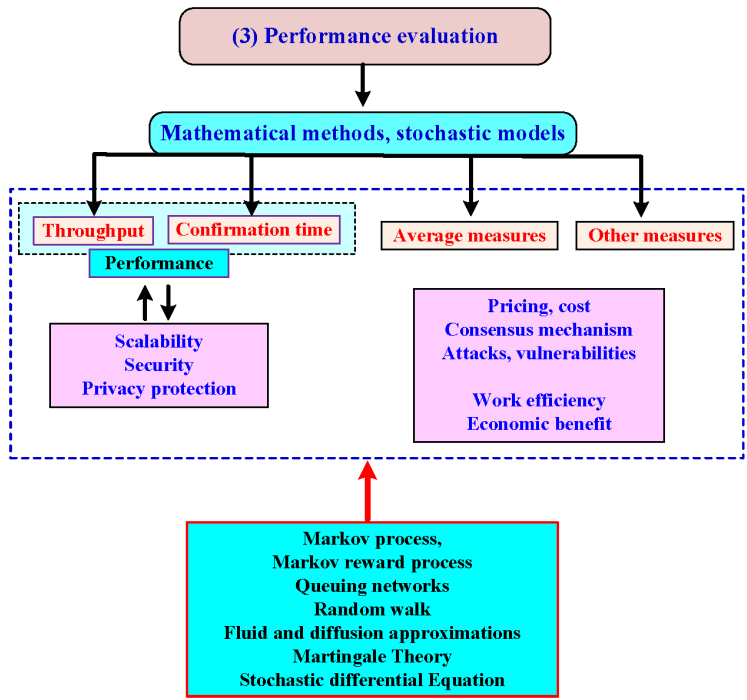} \centering
\caption{Performance modeling and analysis of blockchain systems}
\label{Fig-3}
\end{figure*}

\textbf{(4) to (6) Performance optimization, dynamic decision, and machine learning}

In this class of mathematical modeling and analysis \textbf{(4)}, we need to
optimize the performance measures of a blockchain system by means of linear
programming, nonlinear programming, integer programming, multi-objective
programming and so on.

In this class of mathematical modeling and analysis \textbf{(5)}, we need to
realize optimal control and dynamic decision of a blockchain system by using
the Markov decision processes, sensitivity-based optimization, and
stochastic optimal control. See Figure 4 for more details.

For machine learning \textbf{(6)}, we need to develop machine learning, deep
reinforcement learning, and federated learning. See Figure 4 for more
details.
\begin{figure*}[tbph]
\centering                 \includegraphics[width=10cm]{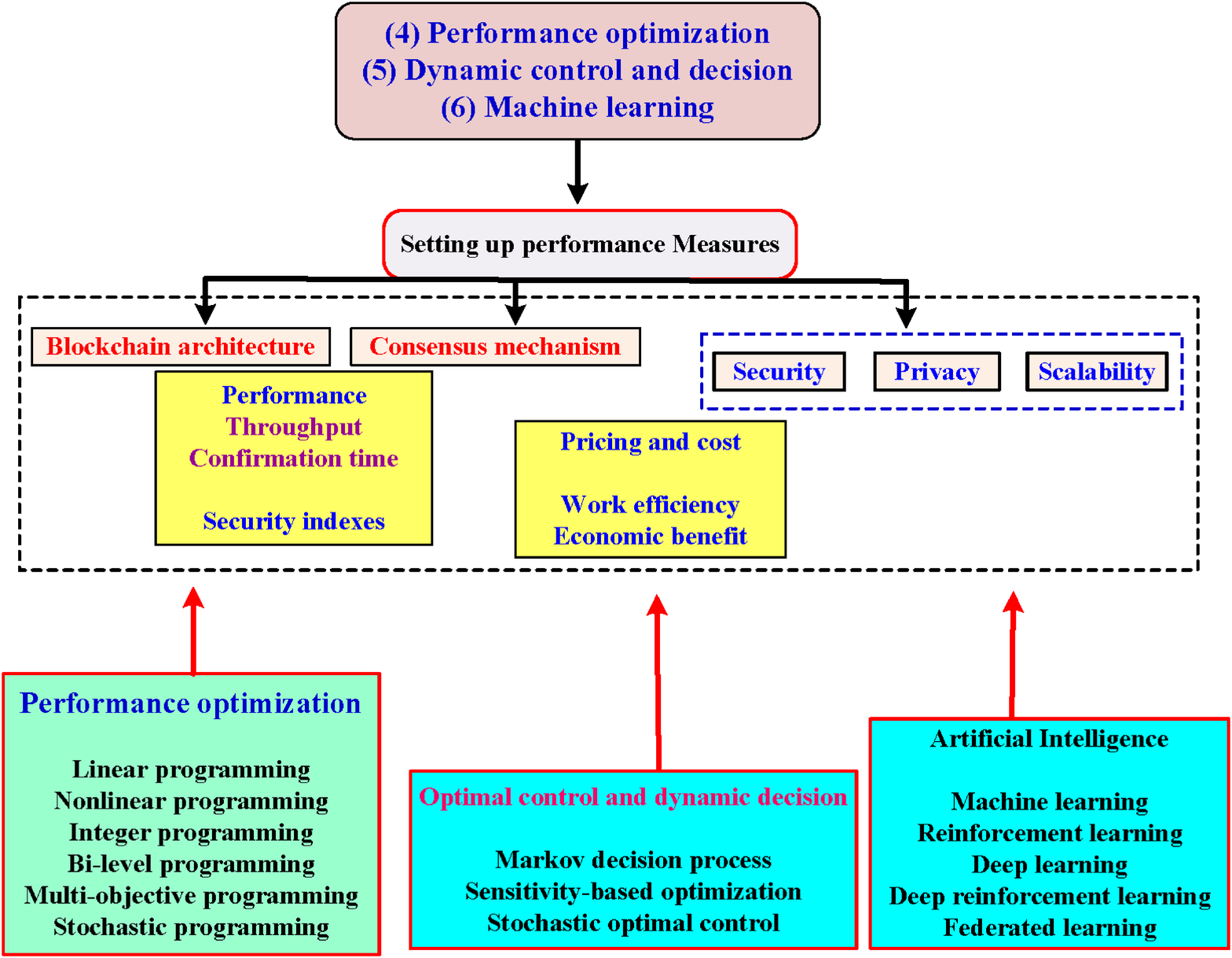} \centering
\caption{Performance optimization, dynamic decision, and machine learning}
\label{Fig-4}
\end{figure*}

\textbf{(7)} \textbf{Blockchain economy and market, and} \textbf{(8)} \textbf{%
blockchain ecology}

For the blockchain economy and market \textbf{(7)} as well as the blockchain ecology \textbf{(8)},
readers may refer to Figures 5 and 6 for a simple introduction, respectively.

\begin{figure*}[tbph]
\centering                 \includegraphics[width=10cm]{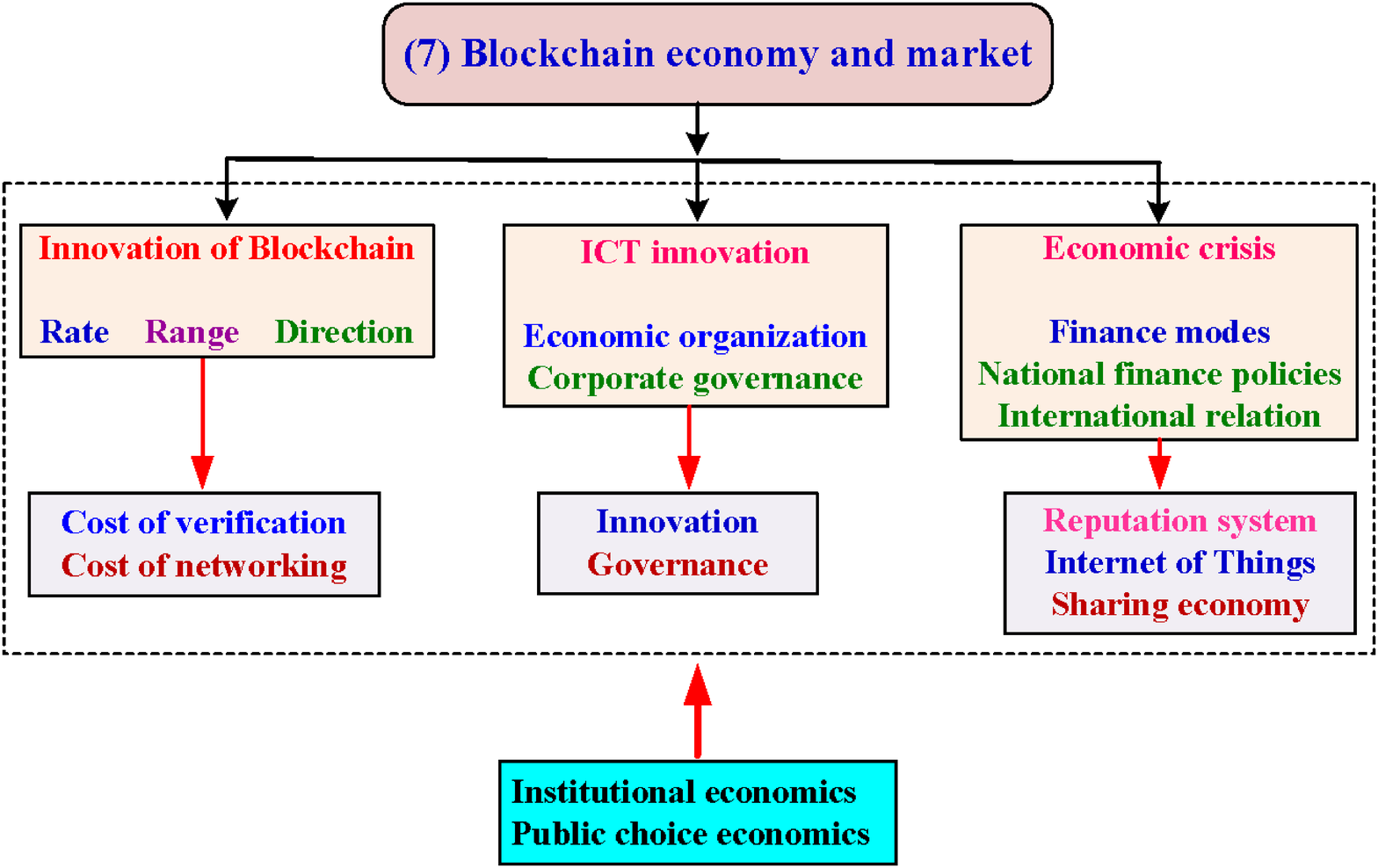} \centering
\caption{The blockchain economy and market}
\label{Fig-5}
\end{figure*}

\begin{figure*}[tbph]
\centering                 \includegraphics[width=10cm]{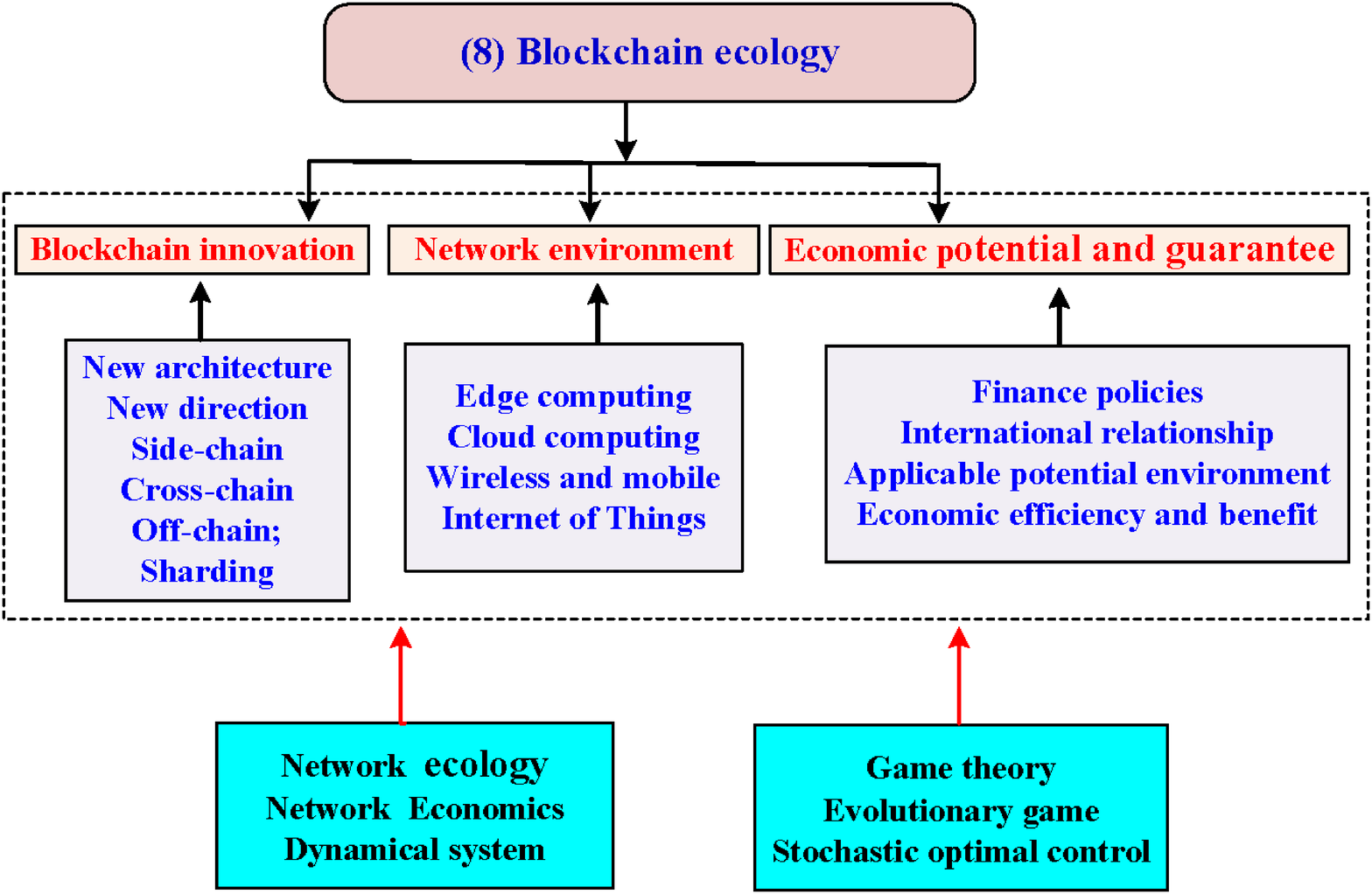} \centering
\caption{The blockchain ecology}
\label{Fig-6}
\end{figure*}

With fast development of blockchain, new blockchain technology continue to
emerge. Thus, performance evaluation, performance optimization, optimal control, and dynamic decision of
blockchain systems become progressively. In particular,
performance modeling and analysis methods have been increasingly lacking and
insufficient up to now, and especially, for dealing with the newly developing blockchain
technology. Blockchain is a hierarchical comprehensive database, and it
operates under a consensus mechanism of distribute systems in a peer-to-peer network. In addition, blockchain
is an interesting and creative combination of multiple computer technologies,
such as encryption techniques, consensus mechanism, security, privacy
protection, and scalability; and wireless, mobility, cloud computing, edge computing, Internet of
Things, and quantum. Therefore, blockchain is always a complicated
stochastic system under a strongly practical environment. In this situation,
performance evaluation, performance optimization, and dynamic decision of blockchain
systems always become interesting and challenging in their theoretical study.

So far, a few survey papers have discussed blockchain technology with a simple
introduction to performance analysis of blockchain systems. See Table \ref{tab:label1} for more details. From Table \ref{tab:label1}, it is easy to
see that those survey papers focused on several key perspectives:
Performance, scalability, security, and privacy protection.

\begin{table}[tbph]
\caption{Survey papers for performance evaluation of blockchain systems}
\label{tab:label1}{\footnotesize 
\centering
\begin{tabular}{|r|p{11.82em}|p{26.045em}|}
\toprule {\textbf{Year}} & \textbf{Surveys or reviews} & \textbf{Research
scope} \\ \hline
2018 & Kim et al. \cite{Kim:2018} & Scalability solutions \\ \hline
2019 & Rouhani and Deters \cite{Rouhani:2019} & Security, performance, and
applications of smart contracts \\ \hline
2019 & Wang et al. \cite{WangYX:2019} & Performance benchmarking tools;
optimization methods \\ \hline
2019 & Zheng et al. \cite{ZhengZS:2019} & Challenges and progresses in
blockchain from a performance and security perspective \\ \hline
2020 & Smetanin et al. \cite{Smetanin:2020} & Effective simulation and modeling
approaches \\ \hline
2020 & Singh et al. \cite{Singh:2020} & Side-chains for improving
scalability, privacy protection, security of blockchain \\ \hline
2020 & Zhou et al. \cite{Zhou:2020} & Scalability of blockchain
\\ \hline
2020 & Yu et al. \cite{YuW:2020} & Sharding for blockchain scalability \\
\hline
2020 & Fan et al. \cite{Fan:2020} & Stochastic models for blockchain systems: Game theory, performance optimization, machine learning, etc. \\ \hline
2021 & Cao et al. \cite{CaoW:2021} & Mathematical models for blockchain such
as stochastic process, game theory, optimization, and machine learning
\\ \hline
2021 & Huang et al. \cite{HuangK:2021} & Performance models, and analysis tools of
blockchain systems or networks \\
\bottomrule
\end{tabular}
}
\end{table}

To open the scope of our survey research on performance evaluation,
performance optimization, optimal control, and dynamic decision of blockchain systems, this paper chooses
a collection of research materials from major scientific journals, international
conferences, and preprint sites including IEEE Xplore, ACM digital library,
Elsevier, SpringerLink, MPDI, arXiv, HAL, and so on. Based on these research materials, we
provide a detailed review and analysis from the literature with respect to research on performance evaluation, performance optimization, and dynamic decision of multiple blockchain
systems, including the consensus mechanism or protocols or algorithms (PoW,
PoS, DPoS, PBFT, DAG and so on), the blockchain types (Bitcoin, Ethereum,
side-chain, cross-chain, off-chain, and so on), and the new network architecture of
blockchain. At the same time, we provide how to set up stochastic models and
to develop effective methods or algorithms for dealing with performance
evaluation, performance optimization, optimal control, and dynamic decision. Note that
such a study of blockchain technology is interesting and challenging in not only the basic theory but
also many practical applications.

Based on the above analysis, we summarize the main contributions of this
paper as follows:

\begin{itemize}
\item[1.] We provide a basic overview for the available
mathematical methods (in particular, stochastic analysis), which greatly support
performance modeling and computation in performance evaluation, performance optimization, optimal control, and dynamic decision.

\item[2.] We provide a clear outline and structure for performance
evaluation and performance optimization of blockchain systems. Important mathematical
methods and techniques include \textbf{(a)}performance evaluation: Markov processes, queueing
theory, Markov reward process, random walk, fluid and diffusion
approximations, martingale theory; and \textbf{(b)}performance optimization: Linear programming,
nonlinear programming, integer programming, and multi-objective programming.

\item[3.] We summarize optimal control and dynamic decision of blockchain
systems by means of, for example, \textbf{(c)} Markov decision process,
sensitivity-based optimization, and stochastic optimal control; and \textbf{%
(d)} machine learning, deep reinforcement learning, and federated learning.
These issues are interesting and challenging with greater potential in future study.
\end{itemize}

The remainder of this paper is organized as follows. Section \ref{sec:models}
reviews the recent literature on the performance evaluation of blockchain
systems by means of the queueing theory, the Markov processes, and the Markov
reward processes. Complementing Section \ref{sec:models}, Section \ref{sec:Further} provides some further methods
for performance evaluation of blockchain systems by using, for example,
the random walks, the fluid approximation, the diffusion approximation, and
the martingale theory. Section \ref{sec:optimization} reviews performance
optimization of blockchain systems by means of the linear programming, the nonlinear
programming, the integer programming, and the multi-objective programming. Section \ref{sec:decision}
focuses on the overview of applications of the Markov decision processes to
find the optimal dynamic strategy of blockchain systems. Section \ref{sec:Learning} summarizes applications of machine learning (e.g., deep
reinforcement learning and federated learning) to performance optimization
and dynamic decision of blockchain systems. Section \ref{sec:concluding}
highlights some concluding remarks.

\section{Performance Evaluation}

\label{sec:models}

In this section, we summarize performance evaluation models of blockchain
systems by means of queueing theory and the Markov process. Note that some other mathematical methods for performance evaluation are
left for the next section.

\subsection{Queueing theory}

Queueing theory is a key mathematical
tool to set up performance measures and performance evaluation of blockchain
systems. Applying queueing theory to performance analysis of blockchain systems is
interesting but challenging since each blockchain system not only is a complicated stochastic system
but also has multiple key factors and a physical structure with different levels. Specifically,
the key factors include \textbf{(1)} transactions arrivals, \textbf{(2)}
transaction fees, \textbf{(3)} block size, \textbf{(4)} network delay,
\textbf{(5)} block generated process (e.g., mining process and voting
process), \textbf{(6)} the pegging process of a block (or a sub-chains),
\textbf{(7)} mining competition among multiple mining pools (e.g., a tree
structure), \textbf{(8)} mining reward, \textbf{(9)} computing power
distribution. The physical structure contains \textbf{(1)}
consensus mechanism (e.g., PoW, PoS, DPoS, PBFT, and DAG), and \textbf{(2)}
scalable structure (e.g., side-chain, cross-chain, and off-chain). The
research objectives of blockchain systems are designed as, for example, \textbf{(a)} performance: Throughput, confirmation
time; \textbf{(b)} security; \textbf{(c)} privacy protection; and \textbf{(d)} scalability. Based on these specific examples, we can see that it is useful and necessary to apply queueing theory to set up
performance models and to analyze performance measures in the study of
blockchain systems.

\begin{figure*}[tbph]
\centering                 \includegraphics[width=11cm]{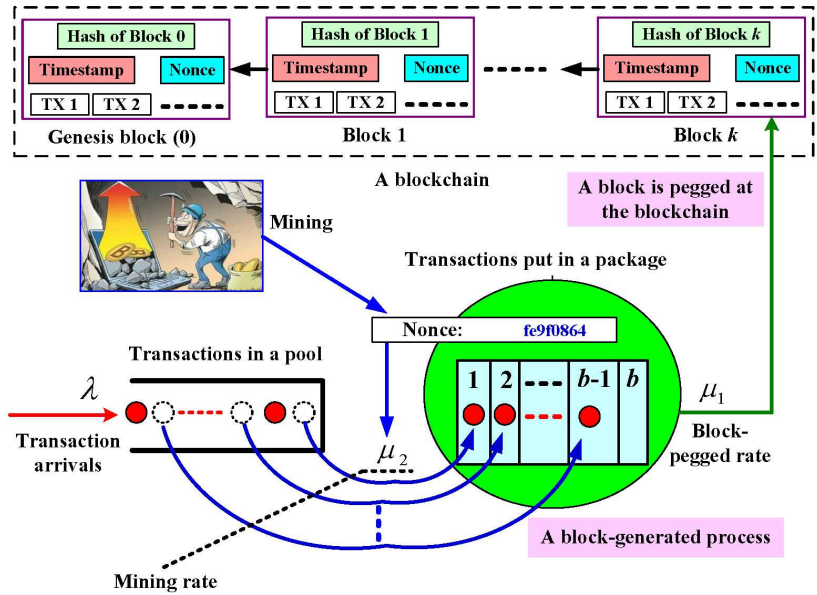} \centering
\caption{A simple physical structure of the PoW blockchain system}
\label{Fig-7}
\end{figure*}

Understanding a blockchain system and its physical structure is not always simple. Li et al. \cite{Li:2018} may be the first one
to provide a simple diagram of the physical structure of the PoW blockchain
system with a miner (or a mining pool). See Figure \ref{Fig-7} for more details.

\begin{figure*}[tbph]
\centering                 \includegraphics[width=9cm]{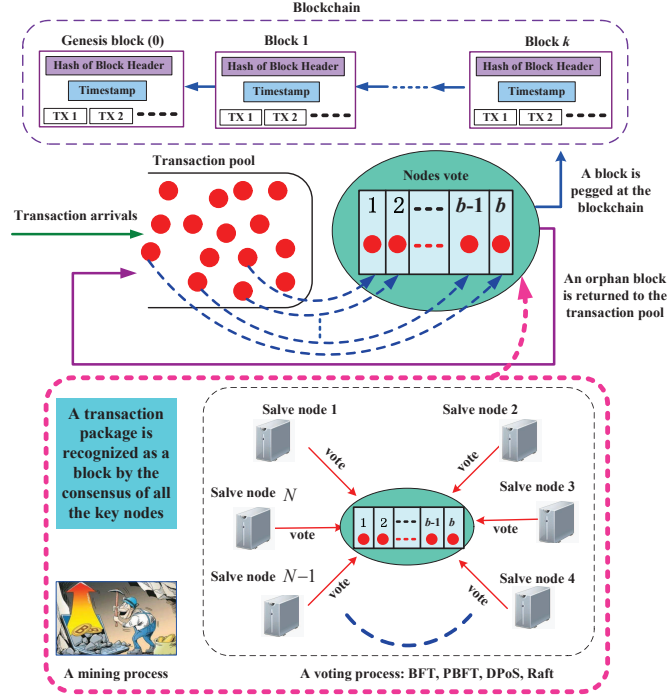} \centering
\caption{A queueing platform of different blockchain systems}
\label{Fig-8}
\end{figure*}

For the other blockchain systems (e.g., PoS, DPoS, BFT, PBFT, and Raft),
Chang et al. \cite{Chang:2022} provided a queueing platform to evaluate their
performance measures once the voting processes are determined by using the Markov modeling technology. Based on this, the first step is to study the voting processes, and the second step is to set up a queueing platform through the voting processes are regarded as the service processes.  See Figure \ref{Fig-8} for more details. In this queueing platform, it first needs to determine
the two random variables: The block-generated time and the
orphan-block-generated time, which can be related to the arrival and
service times in a queueing model $\mathrm{M}\oplus \mathrm{M}^{\mathrm{b}}/\mathrm{M}^{\mathrm{b}}/1$ or $\mathrm{M}\oplus \mathrm{PH}^{\mathrm{b}}/%
\mathrm{PH}^{\mathrm{b}}/1$.

Kawase and Kasahara \cite{Kawa:2017} may be the first to apply queueing theory to study the PoW blockchain system with a miner, and a
further paper by Kasahara and Kawahara \cite{Kasa:2019} considered a
single-server queue with batch service and priority mechanism to analyze the
transaction-confirmation time. Because the block-generation time (note that
it also includes block-pegged time) is a general probability distribution, the
system of differential-difference equations given in the two papers by using the supplementary variable
method will be unsolvable. For this reason, Li et al. \cite{Li:2018}
provided a Markov queue with two stages (the block-generated time and the
block-pegged time) to analyze the PoW blockchain system with a miner. Li et al. \cite{Li:2018}
may be the first paper that clearly describes and expresses the
physical structure with multiple key factors of the PoW blockchain system, as seen in
Figure 7. For the two-stage queue of the PoW blockchain system, the matrix
geometric solution was applied to give a complete solution of this system
such that the performance evaluation of the PoW blockchain system was
established in a simple form and was analyzed by means of a more detailed numerical analysis. In later studies, Li et al. %
\cite{Li:2019} relaxed the model assumptions of Li et al. \cite{Li:2018}
to a more general case that the transaction arrivals are a Markovian arrival
process (MAP), and the block-generated time and the block-pegged time are all of
phase type (PH). Obviously, computing the mean transaction-confirmation time becomes very difficult and complicated due to the complicated blockchain structure, as suggested by Li et al. \cite{Li:2019}.

Kawase and Kasahara \cite{Kawa:2017} and Li et al. \cite{Li:2018} have inspired numerous later strain of literature to use the queueing theory in performance evaluation of blockchain systems. Now, we list some literature as follows:

Geissler et al. \cite{Geiss:2019} neglected the information propagation
delays and assumed the immediate distribution of transactions and blocks to
all the peers in the network. They developed a discrete-time queueing model
that allows performance evaluation of a blockchain system, such as the
transaction waiting time distribution.

Zhao et al. \cite{ZhaoJ:2019} regarded the mining process as a vacation, and
the block-verification process as a service. Specially, they established a
non-exhaustive queueing model with a limited batch service and a possible
zero-transaction service and derived the average number of transactions and
the average confirmation time of a transaction in the blockchain system.

Krieger et al. \cite{Krie:2019} proposed a Markovian non-purging $(n,k)$
fork-join queueing model to analyze the delay time of the
synchronization process among the miners, where a vote-based consensus
procedure is used.

Ahmad et al. \cite{Ahmad:2019} presented an end-to-end blockchain system
for dealing with the audit trail applications, and analyzed the time, space, consensus,
search complexity, and security of this blockchain system by using the queueing
theory.

Mi{\v{s}}i{\'{c}} et al. \cite{Misic:2019} applied the Jackson network model
to the entire network, in which each individual node operates as a priority
M/G/1 queue, and developed an analytical model for analyzing the Bitcoin's
blockchain network.

Frolkova and Mandjes \cite{Frol:2019} proposed a G/M/$\infty $-like Bitcoin
queueing model to consider the propagation delay between two individual users.
Fralix \cite{Fra:2020} provided a further discussion for the infinite-server
queue introduced in Frolkova and Mandjes \cite{Frol:2019}.

Seol et al. \cite{Seol:2020} proposed an embedded Markov chain to analyze a
blockchain system with a specific interest in Ethereum.

He et al. \cite{HeZ:2020} introduced a queueing model with priority to incorporate
the operational feature of blockchain, the interplay between miners and
users, and the security issue associated with the decentralized nature of
the blockchain system.

Gopalan et al. \cite{Gopa:2020} analyzed the stability and scalability of
the DAG-based blockchain system by using queueing networks.

Fang and Liu \cite{Fang:2020} proposed a dynamic mining resources allocation
algorithm (DMRA) to reduce the mining cost in the PoW blockchain networks through using the logical
queueing-based analytical model.

Meng et al. \cite{Meng:2021} proposed a queueing model for studying the
three stages of the consortium blockchain consensus, analyze the consistency
properties of consortium blockchain protocols, and provided performance
evaluation for the main stages of the blockchain consensus.

Sun et al. \cite{Sun:2021} provided a queueing system with three service
stages, which express the three-stage consensus process of the RC-chain and
the building of a new block. By using the queueing model, they obtained
three key performance measures: The average number of transactions in
system, the average transaction confirmation time, and the average
transaction throughput.

Altarawneh et al. \cite{Alta:2021} set up a queueing model to compute the
average waiting time for the victim client transactions, and evaluated the
security and reliability of the blockchain system.

Wilhelmi et al. \cite{Wil:2021} proposed a batch-service queue model for
evaluating the network delay in a blockchain system. Furthermore, they provided
some simulations to assess the performance of the synchronous and
asynchronous mechanisms.

Ricci et al. \cite{Ricci:2019} proposed a framework encompassing machine
learning and a queueing model M/G/1 to identify which transactions will be
confirmed, and characterized the confirmation time of confirmed transactions.

Li et al. \cite{LiY:2018} discussed a queueing game with a non-preemptive
priority of a blockchain system and considered both the miners' mining
rewards and the users' time costs.

Sukhwani et al. \cite{Sukh:2018} presented a performance method of
Hyperledger Fabric v1.0+ by using a stochastic Petri net modeling
(stochastic reward nets) to compute the throughput, utilization, mean
queue length at each peer, and the critical processing stages within a peer.

For ease of reading, we summarize the queueing models of blockchain systems
in Table \ref{tab:label2} .

\begin{table}[tbph]
\caption{The queueing models of blockchain systems}
\label{tab:label2}{\scriptsize \centering
\begin{tabular}{|p{3.045em}|p{3.275em}|p{10.41em}|p{25.135em}|}
\toprule \textbf{Paper} & \textbf{Year} & \textbf{Queue type} & \textbf{%
Research scope} \\ \hline
\cite{Sukh:2018} & 2018 & Petri Nets model & Throughput; utilization; mean
queue length at each peer; critical processing stages within a peer \\ \hline
\cite{LiY:2018} & 2018 & A queueing game & The miners' mining rewards; the users' time cost \\ \hline
\cite{Geiss:2019} & 2019 & GI/G$^X$/1 & Queue size; waiting time of transactions
\\ \hline
\cite{ZhaoJ:2019} & 2019 & M/G$^X\oplus$ G/1 & The average number of
transactions; the average confirmation time of transactions \\ \hline
\cite{Krie:2019} & 2019 & Fork-join queue & The delay performance
of the synchronization process among the miners \\ \hline
\cite{Ahmad:2019} & 2019 & M/D/c & The time, space, consensus, and search
complexity; security \\ \hline
\cite{Misic:2019} & 2019 & Jackson network model; M/G/1 & Probability
distributions of block and transaction distribution time; node response
time; forking probabilities; network partition sizes; duration of ledger's
inconsistency period. \\ \hline
\cite{Ricci:2019} & 2019 & M/G/1 & Identify which transactions will be
confirmed; the confirmation time of confirmed transactions \\
\hline
\cite{Frol:2019} & 2019 & GI/M/$\infty$ & Propagation delay between two
individual users \\ \hline
\cite{Fra:2020} & 2020 & Infinite-server queue & A further study of the infinite-server queue studied in \cite{Frol:2019}; related
infinite-server queues have similar dynamics \\ \hline
\cite{Seol:2020} & 2020 & M$^{X}$/M$^X$/1 & The average number of slots; the
average waiting time per slot; throughput \\ \hline
\cite{HeZ:2020} & 2020 & M/M$^X$/1 with priority & Users' equilibrium
behavior; total fee rate; confirmation latency; system equilibria \\ \hline
\cite{Gopa:2020} & 2020 & Monotone separable queuing models & Stability and
scalability of the DAG network \\ \hline
\cite{Fang:2020} & 2020 & Logical queueing-based analytical model & Mining resources allocation; mining cost; stability \\ \hline
\cite{Meng:2021} & 2021 & M/H$_2$/1; M/M/1; M/Er/1 & The
consistency and security of consortium blockchain protocols \\
\hline
\cite{Alta:2021} & 2021 & M/M/1; M/M/$\infty$ & The average waiting time for
the victim client transactions; security;
reliability \\ \hline
\cite{Sun:2021} & 2021 & Three-phase service queuing process & The average
number of transactions; the average transaction confirmation time;
the average transaction throughput. \\ \hline
\cite{Wil:2021} & 2021 & A novel batch-service queue model & The
learning completion delay of blockchain-enabled federated learning;
performance of synchronous and asynchronous mechanisms \\\hline
\cite{Chang:2022} & 2022 & $\mathrm{M}\oplus \mathrm{M}^{\mathrm{b}}/\mathrm{M}^{\mathrm{b}}/1$ & Throughput of the dynamic PBFT blockchain system; the stationary rate (or probability) that a block is pegged on the blockchain; the stationary rate (or probability) that an orphan block is returned to the transaction pool\\
\bottomrule
\end{tabular}
}
\end{table}

By means of the queueing theory, some papers have conducted research on the
simulation and empirical study of blockchain systems. Important examples
include among which Memon et al. \cite{Memon:2019} and Spirkina et al. \cite%
{Spir:2020} proposed a queueing theory-based simulation model to understand
the performance measures of the blockchain system.

In the queueing models of blockchain systems, Bowden et al. \cite{Bow:2020} is a key
work because the generation time is related to the service time. They
showed that the generation time of a new block has some key statistical
properties, for example, the generation time is non-exponential, and it can
also be affected by many physical factors.

So far, many classes of blockchain systems have still been lacking research
on performance evaluation by using the queueing theory. For example, the PoW
blockchain system with multiple mining pools, the PBFT blockchain system of
dynamic nodes, the DAG-based blockchain systems, the Ethereum, and the
large-scale blockchain systems with either cross-chain, side-chain, or off-chain. Therefore, the queueing models of blockchain systems are always
interesting and challenging in the future study of blockchain technology.

\subsection{Markov processes and Markov reward processes}

In performance evaluation of blockchain systems, Markov processes and Markov
reward processes are two effective mathematical methods. See Li \cite{Li:2010} for a set of Markov models and computational methods by using the RG-factorizations. Note that Markov processes are used to evaluate throughput, confirmation time, and security and
privacy protection of blockchain systems; while the Markov reward processes are applied to analyzing work
efficiency, economic benefits, and cost control of blockchain systems.

For the vulnerability and forked structure of the PoW blockchain systems with two mining pools (honest and dishonest),
Eyal and Sirer \cite{Eyal:2014} proposed a selfish mining strategy for the
competitive mining process between two mining pools, and set up a simple Markov
process with a special reward structure to discuss the competitive behavior
between the two mining pools. By means of an intuitive reward analysis, they
indicated that the selfish miner can win a higher mining reward through
violating the honest agreement in the blockchain system. However, Li et al. %
\cite{Li:2021} showed that the Markov process with rewards given in Eyal and
Sirer \cite{Eyal:2014} is not correct from the ordinary theory of Markov processes.

For a PoW blockchain system with two mining pools (honest and dishonest), Li
et al. \cite{Li:2021} showed the competitive behavior between the two mining
pools by means of Figure 9.

\begin{figure*}[tbph]
\centering                 \includegraphics[width=14cm]{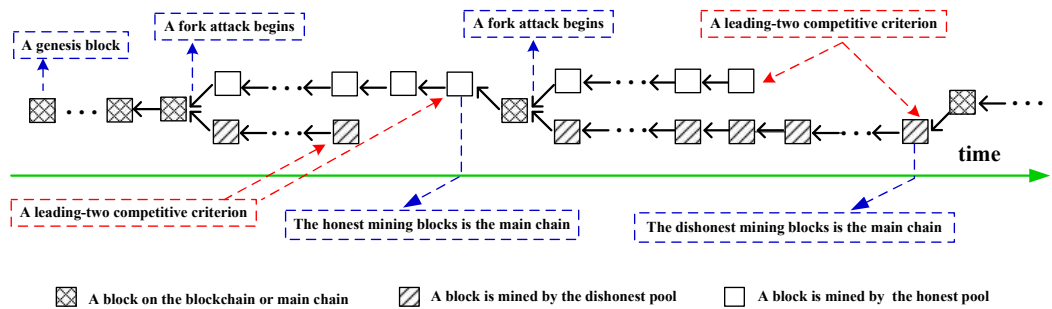} \centering
\caption{The competitive behavior between the two mining pools}
\label{Fig-9}
\end{figure*}

When the two block branches forked at a common tree root, let $I(t)$
and $J(t)$ be the numbers of blocks mined by the honest and dishonest mining
pools at time $t$, respectively. It is seen from Li et al. \cite{Li:2021}
that $\left\{ \left( I(t),J(t)\right) :t\geq 0\right\} $ is a
continuous-time Markov process whose infinitesimal generator is given by
\begin{equation*}
Q=\left(
\begin{array}{ccccccc}
Q_{\widetilde{\mathbf{0}},\widetilde{\mathbf{0}}} & Q_{\widetilde{\mathbf{0}}%
,0} & Q_{\widetilde{\mathbf{0}},1} &  &  &  &  \\
Q_{0,\widetilde{\mathbf{0}}} & Q_{0,0} & Q_{0,1} &  &  &  &  \\
Q_{1,\widetilde{\mathbf{0}}} &  & Q_{1,1} & Q_{1,2} &  &  &  \\
B &  &  & A & C &  &  \\
B &  &  &  & A & C &  \\
\vdots &  &  &  &  & \ddots & \ddots%
\end{array}%
\right) .
\end{equation*}

For the PoW blockchain system with two mining pools (honest and dishonest), G\H{o}bel et al. \cite{Gob:2016} set up a two-dimensional Markov process
with network propagation delay and provided performance evaluation of the
PoW blockchain system. Javier and Fralix \cite{Jav:2020} further discussed
the two-dimensional Markov process given by G\H{o}bel et al. \cite{Gob:2016}
and developed a new computational method. Li et al. \cite{Li:2021} set
up a new two-dimensional pyramidal Markov (reward) process of the blockchain
system, which leads to a novel theoretical framework for performance
evaluation of a PoW blockchain system with adding new random factors by means of a
new class of matrix geometric solutions.

Using the Markov process approach of Eyal and Sirer \cite{Eyal:2014}, Nayak
et al. \cite{Nay:2016} introduced a new type of mining strategy: The
stubborn mining strategy and also established its two extended forms: The
equal-fork stubborn mining strategy and the path stubborn mining strategy.
Further important examples include Wang et al. \cite{WanL:2019} and Liu et al.
\cite{LiuH:2020}. In addition, inspired by the Markov process approach of Eyal and Sirer \cite{Eyal:2014}, the selfish mining strategy was extended to the Ethereum system. Readers can see Grunspan and P{\'{e}}rez-Marco \cite{Grun:2020} and Niu and
Feng \cite{Niu:2019} for more details. Also, the impact of the selfish mining behavior of multiple mining pools on
the blockchain system has also been paid widespread attention, e.g., see Bai et al. \cite{Bai:2019}, Bai et al. \cite{Bai:2021}%
, Chang \cite{Chang:2019}, Liu et al. \cite{Liu:2018}, Marmolejo-Coss\'{\i}o
et al. \cite{Mar:2019} and Xia et al. \cite{Xia:2021}.

From the ordinary theory of Markov processes, we summarize some works that use the Markov processes or Markov reward processes to study other interesting issues of blockchain systems as follows.

Song et al. \cite{Song:2022} provided a Markov process theory for network growth
processes of DAG-based blockchain systems.

Chang et al. \cite{Chang:2022} applied a large-scale Markov process to study the dynamic-PBFT blockchain system.

Carlsten \cite{Carl:2016} applied the Markov process to study the impact
of transaction fees on the selfish mining strategy of the blockchain.

Shi et al. \cite{Shi:2021} developed a new consensus protocol (Proof-of-Age,
PoA) and employed a continuous time Markov chain to show that the consensus
protocol can disincentivize the pooled mining.

Kiffer et al. \cite{Kiffer:2018} set up a Markov-chain to analyze the
consistency properties of blockchain protocols.

Huang et al. \cite{Huang:2019} established a Markov process with an
absorbing state to give performance analysis of the raft consensus algorithm
in private blockchains.

Ma et al. \cite{Ma:2021} established a two-dimensional Markov process to
provide performance evaluation of PBFT blockchain systems.

Srivastava \cite{Sri:2019} computed the transaction confirmation time of
blockchain by using a Markov model.

Li et al. \cite{LiC:2020} established a
Markov process to analyze performance and security of the IoT ledgers with a
directed acyclic graph.

Li et al. \cite{LiC:2021} established the Markov
process to study the block access control mechanism in the wireless
blockchain network.

Piriou and Dumas \cite{Pir:2018} constructed a Markov process to analyze the
blockchain system and developed a simulation model of blockchain technology.

Nguyen et al. \cite{Nguyen:2020} applied the Markov process and deep
reinforcement learning to study the task offloading problem in the mobile
blockchain with privacy protection.

Jofr{\'{e}} et al. \cite{Jof:2021} established a Markov process to study the
convergence rate of blockchain mining games.

Together, these studies outline a critical role of Markov processes and Markov
reward processes in the performance evaluation of blockchain
systems. This would be a potential area for future study.

\section{Further Methods for Performance Evaluation}

\label{sec:Further}

In this section, we summarize further methods for performance evaluation of
blockchain systems, including the random walk, the fluid approximation, the
diffusion approximation, and the martingale theory.

\subsection{The random walk}

The random walk is a key mathematical method in analyzing many stochastic
models, such as queueing systems and information and communication
technology (ICT) systems. See Spitzer \cite{Spi:2001}, Prabhu \cite{Prabhu:1998}, and Xia et al. \cite{Xia:2020} for more details.

Recent, a few papers have studied blockchain systems by
using the random walk, and especially, analyzing the double-spending attacks
of blockchain.

Goffard \cite{Goff:2019} refined a random walk model underlying the
double-spending problem and provided a fraud risk assessment of the
blockchain system.

In contrast with Goffard's model \cite{Goff:2019}, Jang and Lee \cite{Jang:2020} proposed a new random walk model to further study the
probability distribution of catch-up time spent for the fraudulent chain to
catch up with the honest chain, which takes into account the block
confirmation. They discussed the
profitability of the double-spending attacks that manipulate a priori mined
transaction in a blockchain system.

Brown et al. \cite{Brown:2021} studied the duration and probability of
success of a double-spend attack in terms of the random walk.

Grunspan and P\'{e}rez-Marco \cite{Gru:2021} determined the minimal number
of confirmations requested by the recipient such that the double spend
strategy is non-profitable by means of the random walk.

\subsection{The fluid and diffusion approximations}

The fluid and diffusion approximations are two key mathematical methods in
analyzing many stochastic models with general random variables, such as
queueing systems, inventory models, supply chains, and communication
networks. The fluid and diffusion approximations describe a deterministic
process that aims to approximately analyze the evolution of stochastic
processes, that is, they can analyze the evolution of generalized stochastic
processes by using the idea of weak limits. Recently, fluid and
diffusion approximations have been widely used in analyzing of
large-scale complex networks with the tendency of scale expansion, complex
structure, and dynamic state. See Chen and Yao \cite{Chen:2001}, Whitt \cite{Whitt:2002}, Dai et al. \cite{Dai:2012}, B{\"{u}}ke and
Chen \cite{Buke:2017}, Chen and Shanthikumar \cite{Chen:1994} for more
details.

So far, fluid and diffusion approximations have been applied to the analysis
of blockchain systems. Important examples include among which Frolkova and
Mandjes \cite{Frol:2019} developed a Bitcoin-inspired infinite-server model
by means of a random fluid limit. King \cite{King:2021} proposed a fluid
approximation of the random graph model and discussed the related
technologies of shared ledgers and distributed ledgers in blockchain
systems. Ferraro et al. \cite{Fer:2019} studied the stability of unverified
transaction systems in the DAG-based distributed ledgers by means of the
fluid approximation. Koops \cite{Koops:2018} applied the diffusion
approximation to predict the confirmation time of Bitcoin transactions.

There are a few blockchain works that analyze the evolution of generalized
stochastic processes by using the idea of weak limits. For example, Corcino et
al. \cite{Cor:2018} discussed the mean square displacement of fluctuations
of Bitcoin unit prices over time on a daily basis by applying the method of
Brownian motion and Gaussian white noise analysis. Chevallier et al. \cite%
{Cheva:2019} used the L{\'{e}}vy jump diffusion Markov switching model to
study the price fluctuation characteristics of Bitcoin.

For the fluid and diffusion approximations of blockchain systems, it is
interesting and challenging to study the PoW blockchain systems with multiple mining pools. See
Li et al. \cite{Li:2022} for a general tree representation of complicated mining
competition among multiple mining pools. Note that the fluid and diffusion
approximations can also provide performance evaluation of blockchain
systems, thus there exists a great potential and innovation in the future research of many blockchain systems (e.g., PoS, DPoS, PBFT, and DAG).

\subsection{The martingale theory}

The martingale theory not only enriches the contents of probability theory
but also provides a powerful method for studying stochastic processes and stochastic models, and
it is widely applied in economics, networks, decision, and control. Grunspan
and P\'{e}rez-Marco applied the martingale theory to study the profits of miners
under different attacks of blockchain systems since 2018. Using the
martingale theory, the research on common attacks in blockchain systems is
summarized in Table \ref{tab:label22}.

\begin{table}[htbp]
\caption{Research on attacks of blockchain by using martingale theory}
\label{tab:label22}{\scriptsize \centering
\begin{tabular}{|c|p{7.045em}|p{19.82em}|p{15em}|}
\toprule \textbf{Year} & \textbf{Attack type} & \textbf{Research scope} &
\textbf{Method or theory} \\ \hline
2018 & Selfish mining \cite{Gru:2018a} & Expected duration of
attack cycles; the profitability model by using repetition games; improvement of Bitcoin protocol; the miner's attraction to the selfish mining pools &
Martingale theory; Doob stopping time theorem \\ \hline
2018 & Stubborn mining \cite{Gru:2018b} & The profitabilities of stubborn mining strategies & Martingale theory;
Catalan numbers and Catalan distributions \\ \hline
2018 & Trailing mining \cite{Gru:2018c} & The revenue ratio of the
trail stubborn mining strategy in the Bitcoin network; the profitability of other block-withholding strategies & Martingale theory;
classical analysis of hiker problems \\ \hline
2020 & SM; LSM; EFSM and so on \cite{Gru:2020a} & The profitabilities of various mining strategies & Martingale theory; Markov
chains; Dyck words \\ \hline
2020 & SM, intermittent SM and smart mining \cite{Gru:2020b} & The closed forms for the profit lag; the revenue ratio for the strategies
``selfish mining" and ``intermittent selfish mining" & Martingale theory; foundational set-up from previous companion article \\ \hline
2021 & Nakamoto double spend \cite{Gru:2021} & The exact
profitability for Nakamoto double spend strategy; the minimal
number of confirmations to be requested by the recipient such that this
double spend strategy is non-profitable & Martingale theory; glambler ruin;
random walk \\
\bottomrule
\end{tabular}
}
\end{table}

\section{Performance Optimization}

\label{sec:optimization} In this section, we provide an overview for
performance optimization of blockchain systems by using different optimal methods.

Performance optimization is to optimize performance measures of blockchain
systems by means of mathematical programming (e.g., linear programming,
nonlinear programming, integer programming, and multi-objective programming). And it composes four elements: Optimization problem,
optimization variables, objective functions, and restrictive conditions. The
optimization process needs to accomplish such a task: When the restrictive
conditions are satisfied, the optimization variables are adjusted to make
that these objective functions go to a maximum or a minimum.

Performance optimization is necessary and important in the study of
blockchain systems, including design, organization, control, and management
of blockchain systems. Such a study will strongly support the overall development of
theoretical research and practical applications of blockchain technology.

So far, performance optimization of blockchain systems has been studied in
at least three aspects as follows:

\textbf{(1) }From consensus mechanism and network architecture of blockchain
systems, it is interesting to optimize performance (e.g., throughput and
confirmation time), work efficiency, economic benefit; improve scalability,
security, privacy protection and degree of decentralization; and balancing operations
costs and efficiency, and allocation of profits. Important examples include
Lundbaek and D'Iddio \cite{Lund:2016}, Liang \cite{Liang:2018}, Nguyen et
al. \cite{Nguyen:2019}, Wang et al. \cite{WangX:2019}, Saad et al. \cite%
{Saad:2019}, Reddy and Sharma \cite{Reddy:2020}, Leonardos et al. \cite%
{Leo:2020}, Liu et al. \cite{LiuF:2022}, Varma and Maguluri \cite{Varma:2021}%
, and Li et al. \cite{LiZ:2021}.

\textbf{(2) }From some key factors (e.g., operations costs, pricing,
computing power, transaction fee, network delay) of PoW blockchain systems,
it is necessary to consider the optimal strategies of dishonest miners, for
example, how to pack a transaction package from a transaction pool? How to
incentive honest miners to jump into the dishonest mining pool? How to
incentive the dishonest miners to keep mining in a round of competition? How
to maximize miners' economic benefit or work efficiency? Important examples
include Kang et al. \cite{KangX:2018}, Aggarwal et al. \cite{Aggar:2019},
Ramezan et al. \cite{Rame:2020}, and Liu et al. \cite{LiuF:2022}.

\begin{table}[tbph]
\caption{Performance optimization of blockchain systems}
\label{tab:label33}{\tiny 
\centering
\begin{tabular}{|p{6.545em}|p{23.275em}|p{13.41em}|p{6.135em}|}
\toprule \textbf{Proposed for} & \textbf{Optimization scope} & \textbf{%
Optimization factors} & \textbf{Methods} \\ \hline
Governed blockchains \cite{Lund:2016} & Solving the MINLP optimization problems for computing optimal Proof of
Work configuration parameters that trade off potentially conflicting aspects
such as availability, resiliency, security, and cost
& Expected availability; resiliency; security; cost & Mixed integer nonlinear programming \\ \hline
A new system \cite{Liang:2018} & Re-innovating all the core elements of the
blockchain technology to achieve the best balance among scalability,
security and decentralization &  Transaction confirmation
time; information propagation latency & Min-max optimization \\ \hline
Users, miners, and verifiers \cite{KangX:2018} & Considering the tradeoff between the
network delay of block propagation process and offered transaction fee from
the blockchain user to jointly maximize utility of the blockchain user and
individual profit of the miners & Network delay of block propagation
process; offered transaction fee from the blockchain user & Nonlinear programming \\ \hline
A new sharding paradigm \cite{Nguyen:2019} & Proposing OptChain that can
minimize transactions and maintain a temporal balance among shards to
improve the confirmation time and throughput & Confirmation time;
transaction throughput; cross-shard transactions minimization; temporal balancing &
Nonlinear programming \\ \hline
A new dynamic routing solution \cite{WangX:2019} & Proposing a new dynamic routing solution Flash to strike a better tradeoff between path optimality and probing overhead & Payment size;
transaction fees; probing overhead; transaction throughput & Linear programming \\ \hline
Miners \cite{Aggar:2019} & Demonstrating BTC's robust stability, and find
that the implemented design of emergency difficulty adjustment resulted in
maximal miners' profits & Coinbase reward; competition cost reward;
transaction fees; competition cost fees; mining cost; waiting cost;
switching incentive; miners' profits & Mixed integer nonlinear programming \\ \hline
A new form of attacks \cite{Saad:2019} & Studying a new form of attacks that
can be carried out on the memory pools and proposing countermeasures that
optimize the mempool size and help in countering the effects of DDoS attacks
& Attack cost; relay fee; mining fee; memorypool size &
Nonlinear programming \\ \hline
PoW blockchain and blockDAG \cite{Reddy:2020} & Proposing two models to scale the
transaction throughput & Block creation rate; transaction throughput; main
chain block growth rate; propagation delay; risk & Nonlinear programming \\ \hline
PoS protocols \cite{Leo:2020} & Leveraging weighted majority voting rules
that optimize collective decision making to improve the efficiency and
robustness of the consensus mechanism & Validators' voting behavior;
blockchain rewards; collective decision; collective welfare & Mixed integer nonlinear programming\\ \hline
A new pricing mechanism \cite{Riehl:2020} & Presenting a pricing mechanism
that aligns incentives of agents who exchange resources on a decentralized
ledger to greatly increase transaction throughput with minimal loss of
security & Transaction pricing; expected transaction efficiency; block
assembly; transaction throughput; security & Integer linear programming \\ \hline
Miners \cite{Rame:2020} & How should miners pick up transactions from a transaction pool to
minimize the average waiting time per transaction & Average waiting time per
transaction & Mixed integer nonlinear programming \\
\hline
Enterprises and users \cite{ZhouL:2020} & Choosing the most effective
platform from many blockchains to control costs and share data & Technical,
market and popularity indicators; improved global DEA-Malmquist measure &
Nonlinear programming \\ \hline
Lightning and Spider network \cite{Varma:2021} & Setting up a two-sided
queue model and propose a throughput optimal algorithm that stabilizes the
system under any load within the capacity region & Transaction throughput;
arrival rate; capacity region; payment requests & Linear programming \\ \hline
A new protocol \cite{LiZ:2021} & Proposing EntrapNet protocol and optimize
EntrapNet to deal with the fundamental tradeoff between security and
efficiency & Security; efficiency & Nonlinear programming \\ \hline
Protocol designer, users, and miners \cite{LiuF:2022} & Proposing a Fee and
Waiting Tax (FWT) mechanism to improve the incentives for the miners'
participation and blockchain security, and to mitigate blockchain
insufficient fee issue & Storage costs of miners; users' transaction fee;
fee choices and waiting tax for users; transaction waiting time &
Multi-objective programming \\
\bottomrule
\end{tabular}
}
\end{table}

\textbf{(3) }For users or enterprises with pricing, cost, transaction fee,
and platform selection, how to maximize user (or enterprise) utility?
Important examples include Kang et al. \cite{KangX:2018}, Riehl and Ward %
\cite{Riehl:2020}, Zhou et al. \cite{ZhouL:2020}, Varma and Maguluri \cite%
{Varma:2021}, and Liu et al. \cite{LiuF:2022}.

Based on the above analysis, we summarize performance optimization of
blockchain systems in Table \ref{tab:label33}. It is seen from Table \ref{tab:label33} that most of the research on
performance optimization of blockchain systems focuses on discussing the
following issues:

\textbf{(i)} Does there exist a better network architecture or consensus
mechanism such that the blockchain system is more efficient, secure, and
scalable?

\textbf{(ii)} Is there a better application scenario that makes blockchain more consistent and less waste of resources?

\textbf{(iii)} Is there a more effective economic incentive mechanism that makes blockchain more profitable and the cost of operations, verification and
communication lower?

\textbf{(iv)} Is there a better trading platform and a more favorable market
environment that make users in blockchain more usable and more credible
among users?

In a word, performance optimization of blockchain systems is an interesting
and hot frontier research topic, and also there exists a large capacity for research innovation
through discussing broad blockchain systems (e.g., consensus
mechanism and network architectures), for example, cross-chain, side-chain,
off-chain, and interoperability of information and assets among different
chains; data synchronization, data security; pricing, cost, economic
benefit, and work efficiency; scalability, security, and privacy protection.

\section{Markov Decision Processes}

\label{sec:decision} In this section, we apply the
Markov decision processes (MDPs) to the study of blockchain systems and
provide some algorithms for computing the optimal dynamic policy of such a Markov
decision process. For the Markov decision processes, readers may refer to Puterman \cite{Put:2014} and Li et al. \cite{Li:2019a} for more details.

The Markov decision processes are widely applied to deal with the selfish
mining attacks in the PoW blockchain systems because the selfish mining
process needs to choose a series of mining policies to be able to maximize the reward or to minimize the cost.

When a PoW blockchain system has two different miners or mining pools
(honest and dishonest) to compete for a more mining reward, in which the
dishonest miner may adopt different mining policies based on the longest
chain rule. The dishonest miner can control the fork structure of block tree
through releasing some parts of blocks to obtain his maximum benefit.
Accordingly, an interesting topic focuses on how the dishonest miner
finds an optimal mining policy (i.e., how many mined blocks are released in a round of competition).
Important examples include among which Sapirshtein et al. \cite{Sapir:2016},
Sompolinsky and Zohar \cite{Sompo:2016} and Gervais et al. \cite%
{Gervais:2016} introduced four different policies: Adopt, cover, match,
and wait for selfish miners, and they determined the optimal selfish mining
policy.

Zur et al. \cite{Zur:2020} studied the optimal selfish mining policy of
the PoW blockchain system by using the Markov decision process and proposed
a new method to solve the Markov decision process with an average reward
criterion.

Bai et al. \cite{Bai:2021} applied the Markov process to study the PoW
blockchain system with multiple miners and used the Markov decision process
with observable information to find the optimal selfish mining policy for a special case with two different miners.

Li et al. \cite{LiW:2021} discussed the PoW blockchain system by using the
hidden Markov decision process and proposed an improved selfish mining
policy.

Ma and Li \cite{MaL:2021} analyzed the optimal selfish mining policy of
the PoW blockchain system with two mining pools through using the
sensitivity-based optimization theory.

In addition, the Markov decision processes are also applied to deal with other blockchain control
issues as follows:

Niu et al. \cite{NiuW:2021} provided an incentive analysis for the
Bitcoin-NG protocol by using the Markov decision process, and showed that
the Bitcoin-NG protocol can maintain the incentive-compatible mining attacks.

W{\"{u}}st \cite{Wust:2016} used the Markov decision process to study the
data security in the blockchain system.

Chicarino et al. \cite{Chic:2020} discussed the selfish mining inspection
and tracking attacks in the PoW blockchain network by means of the Markov
decision processes.

\section{Machine Learning}

\label{sec:Learning}

In this section, we summarize the applications of machine learning (e.g., deep
reinforcement learning and federated learning) to performance optimization
and dynamic decision of blockchain systems.

Recent, machine learning (e.g., deep reinforcement learning and
federated learning) has been applied to study performance optimization and dynamic
decision of blockchain systems. Since the Markov decision process of a
blockchain system is always more complicated, it is difficult and
challenging to find the optimal policy of the Markov decision process, while
the machine learning can provide an approximate solution for such an optimal
policy. Therefore, it is interesting to develop approximate methods or algorithms to find the optimal policy by using, artificial intelligence,
machine learning, deep reinforcement learning, and federated learning.

\textbf{The survey papers:} Liu et al. \cite{LiuY:2020} provided a survey
for the recent literature that the blockchain technology is analyzed by
means of machine learning and discussed several interesting directions
on this research line. Ekramifard et al. \cite{Ekra:2020} provided a
systematic overview for applying artificial intelligence to the study of
blockchain systems, including the Markov decision process and machine
learning. Chen et al. \cite{Chen:2021} applied machine learning to performance
optimization and dynamic decision of blockchain systems and proposed several
interesting topics for future research. Shafay et al. \cite{Shafay:2022}
reviewed the recent literature on applications of deep reinforcement
learning to develop the blockchain technology.

In what follows, we summarize the recent research on applications of machine learning to the study of blockchain systems from several different
aspects: The mining policy, the mobile-edge computing, and the Internet of
Things or Industrial Internet of Things.

\textbf{The mining policy:} Considering the optimal policy of selfish
mining attacks in Bitcoin as well as the Nash equilibrium in block
withholding attacks, Hou et al. \cite{Hou:2019} proposed a SquirRL framework
to apply deep reinforcement learning to analyze the impact of attacks on
the incentive mechanism of PoW blockchain. Bar-Zur \cite{Bar-Zur:2020} used reinforcement learning to find the
optimal policy for the miners of different sizes through solving a Markov
decision process problem with an average reward criterion. Wang et al. \cite{WangL:2021} applied reinforcement learning (machine
learning) to find the optimal mining policy in the Bitcoin-like
blockchain and designed a new multi-dimensional reinforcement learning
algorithm to solve the mining MDP problem with a non-linear objective
function (rather than a linear objective function in the standard MDP
problems).

When the growth of PoW blockchain is modeled as a Markov decision process, a
learning agent needs to make the optimal decisions over all the states of
Markov environment in every moment. To track the generation of new blocks
and their verification process (i.e., solving the mathematical puzzles), You %
\cite{You:2022} set up the PoW consensus protocol (i.e., solving
mathematical puzzles) through dealing with a reinforcement learning problem.
In this case, the verification and generation of new blocks are
designed as a deep reinforcement learning iterative process.

\textbf{Mobile-edge computing:} Nguyen et al. \cite{Nguyen:2019} applied the
Markov processes and deep reinforcement learning to study the task
offloading problem of mobile blockchain under privacy protection. Qiu et al. %
\cite{Qiu:2019} formulated the online offloading problem as a Markov
decision process and proposed a new model-free deep reinforcement
learning-based online computation offloading approach for the
blockchain-empowered mobile edge computing, in which both the mining tasks
and the data processing tasks are considered. Feng et al. \cite{FengY:2019}
developed a cooperative computation offloading and resource allocation
framework for the blockchain-enabled mobile-edge computing systems and
designed a multi-objective function to maximize the computation rate of
mobile-edge computing systems and the transaction throughput of the
blockchain systems by means of the Markov decision processes.

Asheralieva and Niyato \cite{Ashera:2019} developed a hierarchical learning
framework by means of the Markov decision processes with the service provider
and the miners and studied resource management of edge computing to support
the public blockchain networks. By applying the Markov decision process,
Asheralieva and Niyato \cite{Ashera:2020} presented a novel game-theoretic,
Bayesian reinforcement learning and deep learning framework to represent the
interactions among the miners for the public and consortium blockchains with
mobile edge computing. Yuan et al. \cite{Yuan:2021} applied the Markov decision
processes and deep reinforcement learning to study the sharding
technology for the blockchain-based mobile edge computing.

\textbf{Internet of Things:} Waheed et al. %
\cite{Waheed:2020} provided a summary of the security and privacy
protection of blockchain technology in the Internet of Things by
using machine learning algorithms. Gao et al. \cite{Gao:2020} studied
the task scheduling of the mobile blockchain supporting applications of the
Internet of Things by means of deep reinforcement learning and Markov decision processes.

\textbf{Industrial Internet of Things:} Qiu et al. \cite{Qiu:2018} and Luo et al. \cite{Luo:2020} studied the
blockchain-based software-defined Industrial Internet of Things by means of
a dueling deep Q-learning approach and the Markov decision processes. Yang
et al. \cite{Yang:2020} studied the energy-efficient resource allocation for
the blockchain-enabled Industrial Internet of Things by deep
reinforcement learning and Markov decision processes. Wu et al. \cite%
{Wu:2021} provided a review for the deep reinforcement learning applied to
the blockchain systems in the Industrial Internet of Things.

\section{Concluding Remarks}

\label{sec:concluding}

Since Nakamoto \cite{Nak:2008} proposed Bitcoin in 2008, research on
blockchain has attracted tremendous attention from both theoretical research
and engineering applications. With fast development of blockchain
technology, blockchain has developed many imaginative applicable modes
through a series of innovative combinations among distributed data storage,
point-to-point transmission, consensus mechanisms, encryption techniques,
network and data security, privacy protection, and other computer
technologies. Also, their subversive and imaginative features can further
inspire endless technological innovations of blockchain. Among them, the
most representative technologies, such as timestamp-based chain block
structure, DAG-based network data structure, distributed consensus mechanism, consensus mechanism-based
economic incentives, and flexible and programmable smart contracts, have
increased extremely rich colors to various practical applications. Important
examples include digital economy \cite{Cata:2017}, Fintech \cite{Mori:2016},
cloud services \cite{Gupta:2019}, reputation systems \cite{Dennis:2015},
social security \cite{Tang:2022}, e-commerce supply chain \cite{LiuL:2020}, artificial
intelligence \cite{Hussain:2021}, sharing economy \cite{Hawli:2018}, and
supply chain management \cite{Saberi:2019}.

Performance evaluation, performance optimization, and dynamic decision are one of the
most basic theoretical research of blockchain systems, and they
play a key role in design, control, stability, improvement, and applications
of blockchain systems. So far, some blockchain pitfalls (e.g., low performance and scalability, weak
security and privacy protection, and inconvenient interoperability among
blockchain subsystems) have severely limited a wide range of applications of blockchain technology. To resolve these blockchain pitfalls, a few technologies or methods have been proposed and developed, e.g., see off-chain \cite{Poon:2016}, side-chain and cross-chain \cite{Back:2014}, sharding \cite{Luu:2016}, and DAG \cite{Pop:2016}. However, it is a key to deal with whether these novel technologies and methods can effectively improve
these pitfalls of the blockchain systems, while such an interesting issue is to need to be sufficiently studied by means of some strictly mathematical analysis. On the other hand, it is an interesting
topic to set up some useful mathematic relations among performance,
scalability, security, privacy protection and so forth. Some intuitively understanding examples include among which
increased security will result in low throughput; increased scalability will
result in high throughput; increased security will result in strong
privacy protection. Note that the mathematic relationships can be set up by
means of research on performance evaluation, performance optimization, and dynamic decision of
blockchain systems.

It is easy to understand that practical applications will lead to the
innovation boundary of blockchain technology. That is, blockchain
applications are a main driving force of blockchain technology development. When a
new application of blockchain technology is launched, the interface between
technology and application is not always friendly, the performance and
stability are not always high, and there are also deficiencies in security and
privacy protection. Note that all the necessary improvements or increasing maturity need some plentiful research on performance evaluation, performance optimization,
and dynamic decision of blockchain systems. In addition, for the current blockchain technology, we need to actively create a social
atmosphere and ecological environment for both theoretical research and
practical applications of blockchain. Also, this can powerfully promote deep
integration between the blockchain technology and the key information
technologies (such as artificial intelligence, big data, and the Internet of
Things).

For a large-scale blockchain system or a new blockchain technology, it is
key to find the bottleneck through analyzing vulnerability and fault
tolerance of network architecture by means of some new mathematical theory
and methods developed in research on performance evaluation, performance optimization,
and dynamic decision of blockchain systems. Thus, this motivates us in this paper to provide
a recent systematic overview of performance evaluation, performance optimization, and
dynamic decision of blockchain systems, which involves mathematical modeling
and basic theory of blockchain systems. Important examples include \textbf{(a)} performance evaluation: Markov processes, queuing theory, Markov reward
processes, random walks, fluid and diffusion approximations, and martingale
theory; \textbf{(b)} performance optimization: Linear programming, nonlinear programming, integer
programming, and multi-objective programming; \textbf{(c)} optimal control and
dynamic decision: Markov decision processes, and stochastic optimal control;
and \textbf{(d)} machine learning: Deep reinforcement learning and
federated learning. We believe that the new basic theory with mathematical
methods, algorithms, and simulations discussed in this paper will strongly
support future development and continuous innovation of blockchain
technology.

Based on the above analysis, we believe that there are still many
interesting research directions to be explored, such as smart contract, DAG-based blockchain, cross-chain,
side-chain, off-chain and other network architectures; and some basic or new consensus
protocols. Our future research includes:

-- Developing effective methods to compute and improve performance, stability, and scalability of
blockchain systems.

-- Setting up a mathematical theoretical framework for security and privacy protection of blockchain systems.

-- Providing effective methods to optimize and dynamically control performance, security
and privacy protection of large-scale blockchain systems.

-- Developing machine learning for performance optimization and dynamic decision of blockchain systems.

--Developing a healthy ecological environment and reasonable operations
management in the blockchain community by means of research on
performance evaluation, performance optimization, and dynamic decision of blockchain
systems.

\end{document}